\documentstyle[preprint,aps,axodraw]{revtex}
\tightenlines

\newcommand{\reseteqnum}{\setcounter{equation}{0}}
\newcommand{\nn}{\nonumber}
\newcommand{\eqn}[1]{(\ref{#1})}
\newcommand{\ovl}[1]{\overline{#1}}
\newcommand{\wt}[1]{\widetilde{#1}}
\newcommand{\p}{\partial}
\newcommand{\pslash}{p\kern-1ex /}
\newcommand{\lslash}{l\kern-1ex /}
\newcommand{\kslash}{k\kern-1ex /}
\newcommand{\Dslash}{{\cal D}\kern-1.5ex /}
\newcommand{\bpsi}{\overline{\psi}}
\newcommand{\bq}{\overline{q}}

\newcommand{\vev}[1]{\left\langle #1 \right\rangle}
\begin{document}

%\draft

\title{
\vspace{-3.0cm}
\begin{flushright}  
{\normalsize UTHEP-394}\\
\end{flushright}
One loop renormalization for the axial Ward-Takahashi identity
in Domain-wall QCD}

\author{Sinya Aoki, and Yusuke Taniguchi}

\address{Institute of Physics, University of Tsukuba, 
Tsukuba, Ibaraki 305-8571, Japan 
}

\date{\today}

\maketitle

\begin{abstract}
We calculate one-loop correction to the axial Ward-Takahashi identity
given by Furman and Shamir in domain-wall QCD.
It is shown perturbatively that the renormalized axial Ward-Takahashi
identity is satisfied without fine tuning and
the ``conserved'' axial current receives no renormalization,
giving $Z_A=1$.
This fact will simplify the calculation of the pion decay constant
in numerical simulations since
the decay constant defined by this current needs no lattice renormalization
factor.

\end{abstract}

%11.15Ha : Lattice gauge theory (see also 12.38.G Lattice QCD calculations)
%11.30Rd : Chiral symmetries
%12.38Bx : Perturbative calculations
%12.38Gc : Lattice QCD calculations (see also 11.15.H Lattice gauge theory)

\narrowtext

\section{Introduction}

The lack of chirally invariant fermion formulations is one of the most
uncomfortable points theoretically and practically in lattice QCD.
For example, in the Wilson fermion formulation, which is popularly used
in numerical simulations, the chiral limit can be realized only by
the fine tuning of bare mass parameter, which compensates
the additive quantum correction to the quark mass.

Recently the domain-wall fermion formulation\cite{Shamir93,Shamir95},
which was originally proposed for lattice chiral gauge theories\cite{Kaplan}, 
has been employed in lattice QCD simulations\cite{Blum-Soni}
and has shown its superiority over other formulations:
there seems no need of the fine tuning to realize the chiral limit
while there is no restriction to the number of flavors.
In particular massless mode which presents at the tree level seems stable
against the quantum correction.
This property of the domain-wall fermion seems to suggest the existence
of a kind of axial symmetry.
In Ref.~\cite{Shamir95} Furman and Shamir have defined the axial
transformation and the current in the domain-wall fermion.
They have shown that although their transformation is not a symmetry of the
action, the axial Ward-Takahashi identity is satisfied in Green's
functions between the current operators and physical quark fields
made of boundary fermions.

Although it has already argued in the same reference that
the axial Ward-Takahashi identity seems to hold nonperturbatively,
it is still interesting to see a definite form of
one loop correction to Green's functions in the identity
since renormalized form of the Green's functions is nontrivial.
In this paper we calculate the one loop correction to the Green's functions
with the axial current operators and the quark fields.
It is shown that all the divergence are renormalized into the same Green's
functions without any mixing between these physical operators and
unphysical heavy fermion operators.
The renormalized axial Ward-Takahashi identity is satisfied without any
fine tuning and
the renormalization factor of the axial current becomes unity, $Z_A=1$.

This paper is organized as follows.
In Sec.~II we introduce the action of domain-wall QCD.
The axial transformation and its Ward-Takahashi identity is given
according to Ref.~\cite{Shamir95}.
In Sec.~III we calculate one loop correction to the Ward-Takahashi identity.
The Feynman rules relevant to our calculations are defined in our previous
paper\cite{AIKT98} and are briefly given in the appendix.
Our conclusion is given in Sec.~IV.

In this paper we set the lattice spacing $a = 1$ and take $SU(N_c)$ gauge
group with the gauge coupling $g$.
%and the second Casimir 
%$C_F = \displaystyle \frac{N_c^2-1}{2N_c}$.
%$C_F = (N_c^2-1)/2N_c$.

\section{Axial current and Ward-Takahashi identity}

The domain-wall fermion is a 4+1 dimensional Wilson fermion with
a ``mass term''  which depends on the coordinate in the extra dimension.
For the explicit form of the ``mass term'' 
we adopt the Shamir's one\cite{Shamir93} and the corresponding action
becomes
%and the quark action with the current quark mass $m$ becomes
\begin{eqnarray}
S_{\rm DW} &=&
\sum_{n} \sum_{s=1}^N \Biggl[ \frac{1}{2} \sum_\mu
\left( \bpsi(n)_s (-r+\gamma_\mu) U_\mu(n) \psi(n+\mu)_s
+ \bpsi(n)_s (-r-\gamma_\mu) U_\mu^\dagger(n-\mu) \psi(n-\mu)_s \right)
\nn\\&&
+ \frac{1}{2}
\left( \bpsi(n)_s (1+\gamma_5) \psi(n)_{s+1}
+ \bpsi(n)_s (1-\gamma_5) \psi(n)_{s-1} \right)
+ (M-1+4r) \bpsi(n)_s \psi(n)_s \Biggr]
\nn\\&+&
 m \sum_n \left( \bpsi(n)_{N} P_{+} \psi(n)_{1}
+ \bpsi(n)_{1} P_{-} \psi(n)_{N} \right),
\end{eqnarray}
where $n$ is a 4 dimensional space-time coordinate and $s$ is an extra 
fifth dimensional or flavor index,
the Dirac ``mass'' $M$ is a parameter of the theory
which we set $0 \le M \le 2$ to realize the massless fermion at tree level,
$m$ is a physical quark mass,
and the Wilson parameter is set to $r=-1$.
It is important to notice that we have boundaries for the flavor space;
$1 \le s \le N$.
$P_\pm$ is a projection operator $P_\pm = (1 \pm \gamma_5)/2$.

The remarkable property of the domain-wall fermion is that there exists
a massless fermion mode in the $N\to\infty$ limit at small momentum
at $m=0$.
This massless fermion stays near the boundaries of
the flavor space with the left and the right mode on the opposite side.
At tree level the massless mode $\chi_0$ is given explicitly in
zero momentum limit as
\begin{eqnarray}
\chi_0 = \sqrt{1-w_0^2}
\left( P_+ w_0^{s-1} \psi_s + P_- w_0^{N-s} \psi_s \right),
\end{eqnarray}
where $w_0 = 1-M$.
Although the zero mode is stable against quantum
correction\cite{Aoki-Taniguchi}, the damping factor $w_0$ is renormalized
due to the additive correction to the Dirac mass $M$.
Thus
in numerical simulations
it is more convenient to use the interpolating ``physical'' quark field
defined by the boundary fermions:
\begin{eqnarray}
q(n) = P_+ \psi(n)_1 + P_- \psi(n)_N,
\nn \\
\ovl{q}(n) = \bpsi(n)_N P_+ + \bpsi(n)_1 P_-.
\label{eq:quark}
\end{eqnarray}

The axial transformation is defined to rotate the right and left mode
of the Dirac fermion oppositely.
In the domain-wall fermion the massless mode is localized at the boundary
of the flavor space with the right and left mode in the different side.
The axial transformation of the domain-wall fermion is given by the
flavor dependent vector like rotation of the fermion field
which transform the different boundary fermions with opposite charge
\cite{Shamir95},
\begin{eqnarray}
\delta^a \psi(n)_s = \epsilon \left( \frac{N}{2}+\frac{1}{2}-s \right)
 i \lambda^a \psi(n)_s,
\\
\delta^a \bpsi(n)_s = -\epsilon \left( \frac{N}{2}+\frac{1}{2}-s \right)
i \bpsi(n)_s \lambda^a,
\end{eqnarray}
where $\lambda^a$ is a SU($N_f$) generator of the rotation where $N_f$ is
a number of ``real'' quark flavors, and
$\epsilon(s)$ is a step function,
\begin{equation}
\epsilon(s) =
\left\{
\begin{array}{rr}
 1, & s \ge 0,\\
-1, & s < 0  .\\
\end{array}
\right.
\end{equation}
This transformation acts on the quark field as a usual axial rotation,
\begin{eqnarray}
\delta^a q(n) = i \lambda^a \gamma_5 q(n) ,\quad
\delta^a \bq(n) = i \bq(n) \gamma_5 \lambda^a.
\end{eqnarray}
The corresponding axial current is given by the sum of the conserved
vector current in the Wilson fermion over the flavor index,
\begin{eqnarray}
A^a_\mu(n) &=& 
\sum_{s} \epsilon \left(\frac{N}{2}+\frac{1}{2}-s\right) J^a_\mu(n,s),
\\
J_\mu^a(n,s) &=&
\frac{1}{2} \Biggl(
 \bpsi(n)_s \left( 1+\gamma_\mu \right) U_\mu(n) \lambda^a \psi(n+\mu)_s
-\bpsi(n+\mu)_s \left( 1-\gamma_\mu \right) U_\mu^\dagger(n) \lambda^a
 \psi(n)_s
\Biggr).
\end{eqnarray}

Unfortunately this transformation is not a symmetry of the
action and therefore the axial current is not conserved.
The divergence of the axial current becomes
\begin{eqnarray}
\nabla_\mu A_\mu^a(n)
= 2m \bq(n) \lambda^a \gamma_5 q(n) - 2 X^a(n),
\end{eqnarray}
where $\nabla_\mu$ is a backward derivative and 
$X^a$ is the explicit breaking term characteristic for the
domain-wall fermion,
\begin{eqnarray}
X^a(n) = 
\bpsi(n)_\frac{N}{2} P_+ \lambda^a \psi(n)_{\frac{N}{2}+1}
-\bpsi(n)_{\frac{N}{2}+1} P_- \lambda^a \psi(n)_\frac{N}{2}.
\end{eqnarray}
In Ref.~\cite{Shamir95} it is argued that the proper axial Ward-Takahashi
identity is realized in $N \to \infty$ limit
if we consider the Green's functions with physical quark operators,
\begin{eqnarray}
\nabla_\mu \vev{A_\mu^a(n) {\cal O}}
- 2m \vev{\bq(n) \lambda^a \gamma_5 q(n) {\cal O}}
+i \vev{\delta^a {\cal O}} = 0,
\end{eqnarray}
where operator ${\cal O}$ is made of the quark field \eqn{eq:quark}.
The reason for this is that
the explicit breaking term $X^a$ and the physical operator
${\cal O}$ is separated by $N/2$ in the flavor space and the Green's
function $\vev{X^a(n) {\cal O}}$ is suppressed by a factor $e^{-\alpha N/2}$,
where $\alpha$ is some positive number.

In this paper we set ${\cal O}=q(y) \bq(z)$ and calculate the one loop
correction to the Ward-Takahashi identity.

\reseteqnum
\section{One loop correction to Ward-Takahashi identity}

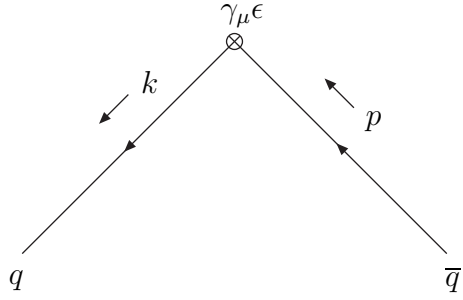
\begin{figure}
\begin{center}\begin{picture}(200,100)(0,0)
\ArrowLine(180,10)(100,90)
\ArrowLine(100,90)(20,10)
\BCirc(100,90){3}
\Line(98,88)(102,92)
\Line(98,92)(102,88)
\Text(150,60)[l]{$p$}\LongArrow(145,65)(135,75)
\Text(65,75)[l]{$k$}\LongArrow(60,70)(50,60)
\Text(95,100)[l]{$\gamma_\mu \epsilon$}
\Text(15,0)[l]{$q$}
\Text(180,0)[l]{$\ovl{q}$}
\end{picture}
\end{center}
\caption{Tree level Green's function with axial current vertex.}
\label{fig:tree}
\end{figure}

In this section we consider the following Ward-Takahashi identity with
${\cal O}=q(y) \bq(z)$,
\begin{eqnarray}
0&=&
\nabla_\mu \vev{ A_\mu^a(n) q(y) \ovl{q}(z)}
-2m \vev{\left(\ovl{q}(n) \lambda^a \gamma_5 q(n)\right) q(y) \ovl{q}(z)}
+2 \vev{X^a(n) q(y) \ovl{q}(z)}
\nn\\&&
+\delta_{n,y} \lambda^a \gamma_5 \vev{q(n) \ovl{q}(z)}
+\delta_{n,z} \vev{q(y) \ovl{q}(n)} \gamma_5 \lambda^a.
\label{eqn:WT-id}
\end{eqnarray}
Here the axial current for domain-wall quarks is given
by a sum of the conserved vector current of the Wilson fermion system.
%over the flavor index.
One may wonder how this ``vector'' current turns out to be the continuum
axial vector current $\bpsi \gamma_\mu \gamma_5 \psi$
in the $a\to0$ limit.
To see this we first study the continuum form of the Green's function
$\vev{A_\mu^a(n) q(y) \bq(z)}$ at tree level and in $N\to\infty$ limit
before the one loop calculation.
In momentum space the tree level Green's function is given by
Fig.~\ref{fig:tree} and has the following form in the continuum limit,
\begin{eqnarray}
\vev{A_\mu^a(q) q(k) \bq(p)}_0 =
\vev{q(-k) \bpsi(k,s)} \gamma_\mu \epsilon(s,t) 
\vev{\psi(-p,t) \ovl{q}(p)} ,
\end{eqnarray}
where $\epsilon(s,t)$ is a diagonal matrix defined by
\begin{eqnarray}
\epsilon(s,t) = \epsilon \left(\frac{N}{2}+\frac{1}{2}-s\right) \delta_{s,t}
\end{eqnarray}
and $\vev{q(-k) \bpsi(k,s)}$, $\vev{\psi(-p,t) \ovl{q}(p)}$ are the
external quark line propagator\cite{AIKT98}.
A summation over the same flavor index is taken implicitly in this paper.
Hereafter we will omit the SU($N_f$) generator $\lambda^a$ for simplicity.

We first notice the fact that the quark external line propagator is
written as a product of the physical quark propagator and some damping
factor in the continuum,
\begin{eqnarray}
&&
\vev{q(-p) \bpsi(p,s)}
\nn\\&&\quad
\to
\frac{1-w_0^2}{i\pslash + (1-w_0^2)m}
\Biggl[
\left( w_0^{(N-s)} P_+ + w_0^{(s-1)} P_- \right)
-\frac{w_0}{1-w_0^2} i\pslash 
\left( w_0^{(s-1)} P_+ + w_0^{(N-s)} P_- \right) \Biggr],
\label{eqn:ext-line1}
\\&&
\vev{\psi(-p,s) \ovl{q}(p)} 
\nn\\&&\quad
\to
\Biggl[
 \left( w_0^{(N-s)} P_- + w_0^{(s-1)} P_+ \right)
-\left( w_0^{(s-1)} P_- + w_0^{(N-s)} P_+ \right)
\frac{w_0}{1-w_0^2} i\pslash \Biggr]
\frac{1-w_0^2}{i\pslash + (1-w_0^2)m}.
\label{eqn:ext-line2}
\end{eqnarray}
We have expanded the propagator to the next to leading order in the
quark external momentum and mass, both of which
give the leading order contribution in the one loop corrections.
Now we can use the formula
\begin{eqnarray}
&&
\left( w_0^{N-s} P_- + w_0^{s-1} P_+ \right) \epsilon(s,t)
=\left(-w_0^{N-s} P_- + w_0^{s-1} P_+ \right)
=\gamma_5 \left(w_0^{N-s} P_- + w_0^{s-1} P_+ \right),
\\&&
\left( w_0^{N-s} P_+ + w_0^{s-1} P_- \right) \epsilon(s,t)
=\left(-w_0^{N-s} P_+ + w_0^{s-1} P_- \right)
=-\gamma_5 \left(w_0^{N-s} P_+ + w_0^{s-1} P_- \right),
\label{eqn:formula}
\end{eqnarray}
which is valid at $N\to\infty$ limit neglecting $w_0^{N/2}$ terms.
For this formula $\gamma_5$ factor appears and the leading order term of
Green's function becomes
\begin{eqnarray}
\vev{A_\mu^a(q) q(k) \bq(p)}_0 = 
\frac{1-w_0^2}{i\kslash + (1-w_0^2)m}
\frac{1}{1-w_0^2} \gamma_\mu \gamma_5
\frac{1-w_0^2}{i\pslash + (1-w_0^2)m}.
\end{eqnarray}
This is a proper form of the Green's function in the continuum
except for the overall factor $1-w_0^2$.

The tree level Green's function with the explicit breaking term vanishes
in the continuum.
%The Green's function is given by
\begin{eqnarray}
\vev{X^a(q) q(k) \bq(p)}_0 &=& \sum_{s,t}
\vev{q(-k) \bpsi(k,s)} 
\left( P_+ \delta_{s,\frac{N}{2}} \delta_{t,\frac{N}{2}+1}
- P_- \delta_{s,\frac{N}{2}+1} \delta_{t,\frac{N}{2}} \right)
\nn\\&&\times
\vev{\psi(-p,t) \ovl{q}(p)} .
\end{eqnarray}
It is easily shown that this is suppressed by $w_0^{N/2}$ in $N\to\infty$
limit because of the damping factor in
\eqn{eqn:ext-line1} and \eqn{eqn:ext-line2}.

Now we calculate the one loop corrections to each terms in the
Ward-Takahashi identity \eqn{eqn:WT-id}.
We set the external quark momentum and the quark mass to the physical
scale, which is much smaller than the cut-off.
We make an expansion in these variables and pick up the leading order terms
which are relevant for renormalization.
The next to leading terms are higher order errors in the lattice spacing.

The one loop correction to the quark propagator and the pseudo scalar
density vertex is already evaluated\cite{AIKT98,Aoki-Taniguchi,Wingate}.
The one loop level full quark propagator and 
the Green's function with pseudo scalar density become
\begin{eqnarray}
&&
\vev{q(-p) \bq(p)}_1^{\rm full} =
\frac{(1-w_0^2)Z_w Z_2}{i\pslash + (1-w_0^2)Z_w Z_m^{-1} m},
\\&&
\vev{(\ovl{q} \gamma_5 q) q(k) \ovl{q}(p)}_1^{\rm full} =
\frac{1}{i\kslash + (1-w_0^2)Z_w Z_m^{-1} m}
\left(1-w_0^2\right) Z_w Z_P^{\rm lat} \gamma_5
\frac{\left(1-w_0^2\right) Z_w Z_2}{i\pslash + (1-w_0^2)Z_w Z_m^{-1} m},
\nn\\
\end{eqnarray}
where $Z_2$, $Z_m$, $Z_P^{\rm lat}$ and $Z_w$ are lattice renormalization 
factors for
the quark wave function, the quark mass, the pseudoscalar density operator,
and the overall factor $w_0$, respectively. ($\lambda^a$ is still omitted.)
Their definitions are given in Ref.~\cite{AIKT98}.

\begin{figure}
\begin{center}
\begin{picture}(200,100)(0,0)
\ArrowLine(180,10)(100,90)
\ArrowLine(40,30)(20,10)
\ArrowLine(80,70)(40,30)
\ArrowLine(100,90)(80,70)
\GlueArc(48,62)(30,-100,10){-5}{6}
\Vertex(40,30){3}
\Vertex(80,70){3}
\BCirc(100,90){3}
\Line(98,88)(102,92)
\Line(98,92)(102,88)
\Text(90,30)[c]{$l-k$}\LongArrow(85,35)(95,45)
\Text(60,72)[l]{$l$}\LongArrow(55,65)(45,55)
\Text(98,100)[c]{$A_0, X$}
\Text(195,20)[l]{$p$}\LongArrow(190,15)(180,25)
\Text(5,20)[r]{$k$}\LongArrow(20,25)(10,15)
\Text(15,0)[l]{$q$}
\Text(180,0)[l]{$\ovl{q}$}
\end{picture}
\end{center}
\begin{center}
\begin{picture}(200,100)(0,0)
\ArrowLine(180,10)(100,90)
\ArrowLine(60,50)(20,10)
\ArrowLine(100,90)(60,50)
\GlueArc(70,40)(12,-225,135){-5}{6}
\Vertex(60,50){3}
\BCirc(100,90){3}
\Line(98,88)(102,92)
\Line(98,92)(102,88)
\Text(98,100)[c]{$A_0, X$}
\Text(195,20)[l]{$p$}\LongArrow(190,15)(180,25)
\Text(5,20)[r]{$k$}\LongArrow(20,25)(10,15)
\Text(15,0)[l]{$q$}
\Text(180,0)[l]{$\ovl{q}$}
\end{picture}
\end{center}
\begin{center}
\begin{picture}(200,100)(0,0)
\ArrowLine(180,10)(160,30)
\ArrowLine(160,30)(120,70)
\ArrowLine(120,70)(100,90)
\ArrowLine(100,90)(20,10)
\GlueArc(152,62)(30,170,280){-5}{6}
\Vertex(160,30){3}
\Vertex(120,70){3}
\BCirc(100,90){3}
\Line(98,88)(102,92)
\Line(98,92)(102,88)
\Text(150,60)[l]{$l$}\LongArrow(145,65)(135,75)
\Text(110,30)[c]{$l-p$}\LongArrow(105,45)(115,35)
\Text(98,100)[c]{$A_0, X$}
\Text(195,20)[l]{$p$}\LongArrow(190,15)(180,25)
\Text(5,20)[r]{$k$}\LongArrow(20,25)(10,15)
\Text(15,0)[l]{$q$}
\Text(180,0)[l]{$\ovl{q}$}
\end{picture}
\end{center}
\begin{center}
\begin{picture}(200,100)(0,0)
\ArrowLine(180,10)(140,50)
\ArrowLine(140,50)(100,90)
\ArrowLine(100,90)(20,10)
\GlueArc(130,40)(12,-315,45){-5}{6}
\Vertex(140,50){3}
\BCirc(100,90){3}
\Line(98,88)(102,92)
\Line(98,92)(102,88)
\Text(98,100)[c]{$A_0, X$}
\Text(195,20)[l]{$p$}\LongArrow(190,15)(180,25)
\Text(5,20)[r]{$k$}\LongArrow(20,25)(10,15)
\Text(15,0)[l]{$q$}
\Text(180,0)[l]{$\ovl{q}$}
\end{picture}
\end{center}
\caption{One loop diagrams which contribute to the quark wave function.}
\label{fig:wave}
\end{figure}
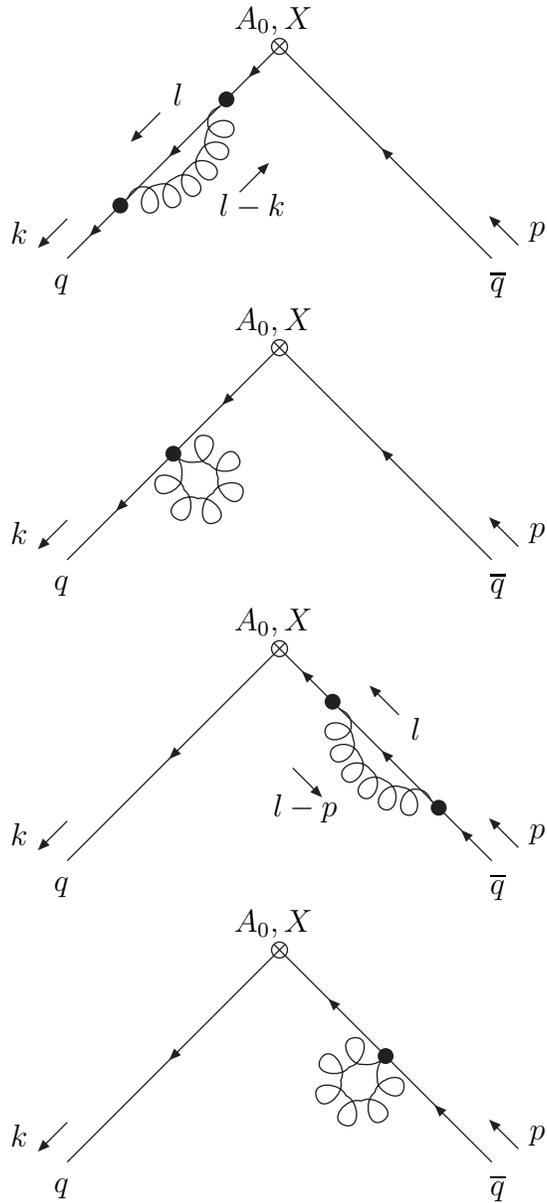

A one loop contribution to the Green's function
$\vev{A_\mu^a(q) q(k) \bq(p)}$
is given by the diagrams in Figs.~\ref{fig:vertex} and \ref{fig:wave}
with vertices $A_i \cdot \epsilon$ which come from the axial vector current.
The one loop diagrams in Fig.~\ref{fig:vertex} contribute to the current
vertex correction,
\begin{eqnarray}
%\nabla_\mu
\vev{A_\mu^a(q) q(k) \bq(p)}_1^{\rm vertex}
&=&
%i(-k_\mu + p_\mu) 
\frac{1-w_0^2}{i\kslash + (1-w_0^2)m}
\frac{T_A^{(1)}}{1-w_0^2} \gamma_\mu \gamma_5
\frac{1-w_0^2}{i\pslash + (1-w_0^2)m},
\nn\\
\end{eqnarray}
where $T_A^{(1)}$ is obtained by multiplying the vertex correction 
with the external propagator damping factor,
\begin{eqnarray}
\frac{T_A^{(1)}}{1-w_0^2} \gamma_\mu \gamma_5
&=&
\left( w_0^{(N-s)} P_+ + w_0^{(s-1)} P_- \right)
\left( \wt{\Gamma}^{(0)}_\mu + \wt{\Gamma}^{(1)}_\mu + \wt{\Gamma}^{(2)}_\mu
 + \wt{\Gamma}^{(3)}_\mu \right)(s,t)
\nn\\&&\times
\left( w_0^{(N-t)} P_- + w_0^{(t-1)} P_+ \right).
\label{eqn:TA}
\end{eqnarray}
$\wt{\Gamma}^{(i)}$'s are contributions from each diagrams represented by
the massless fermion propagator $S_F$, fermion-gluon vertex $V_{1\mu}$,
gluon propagator $G_{\mu \nu}^{ab}$ and axial current vertex
 $A^{(i)}_\mu(l)$ given in the appendix,
\begin{eqnarray}
&&
\wt{\Gamma}^{(0)}_\mu(s,t)=\int_{-\pi}^\pi \frac{d^4 l}{(2\pi)^4}
 \sum_{\nu \rho}
V_{1\nu}^a(0,l) S_F(l)_{s,u} A^{(0)}_\mu(l) \epsilon(u,u') S_F(l)_{u',t}
V_{1\rho}^b(-l,0) G_{\nu \rho}^{ab}(l),
\label{eqn:vertices0}
\\&&
\wt{\Gamma}^{(1)}_\mu(s,t)=\int_{-\pi}^\pi \frac{d^4 l}{(2\pi)^4} \sum_\nu
V_{1\nu}^a(0,l) S_F(l)_{s,u} A^{(1)b}_\mu(l) G_{\nu \mu}^{ab}(l)
\epsilon(u,t),
\\&&
\wt{\Gamma}^{(2)}_\mu(s,t)=\int_{-\pi}^\pi \frac{d^4 l}{(2\pi)^4} \sum_\nu
\epsilon(s,u) A^{(1)a}_\mu(l) S_F(l)_{u,t} V_{1\nu}^b(-l,0)
G_{\mu \nu}^{ab}(l),
\\&&
\wt{\Gamma}^{(3)}_\mu(s,t)=\int_{-\pi}^\pi \frac{d^4 l}{(2\pi)^4}
A^{(2)ab}_\mu G_{\mu \mu}^{ab}(l) \epsilon(s,t).
\label{eqn:vertices3}
\end{eqnarray}
Here we have set the external quark momentum $k_\mu$, $p_\mu$ and
the quark mass $m$ to be zero in the loop because the terms dependent on
them are ${\cal O}(a)$ and do not affect the renormalization.

We can easily see that $\epsilon(s,t)$ 
%acts directly on the damping factors
is multiplied directly by the damping factors
in \eqn{eqn:TA} for $\wt{\Gamma}^{(1)}_\mu$, $\wt{\Gamma}^{(2)}_\mu$ and
 $\wt{\Gamma}^{(3)}_\mu$ and we can use the formula \eqn{eqn:formula}.
For $\wt{\Gamma}^{(0)}_\mu$ we have a relation for massless fermion
propagator,
\begin{eqnarray}
&&
S_F(l)_{t,s} \left(w_0^{(N-s)} P_+ + w_0^{(s-1)} P_-\right)
\nn\\&&=
\left[ P_{+} \left(a_1 w_0^{t-1} + a_2 e^{-\alpha(t-1)} \right)
     + P_{-} \left(b_1 w_0^{N-t} + b_2 e^{-\alpha(N-t)} \right) \right]
(-i \gamma_\mu \sin p_\mu)
\nn\\&&
+\left[P_{+} \left(c_1 w_0^{N-t} + c_2 e^{-\alpha(N-t)} \right)
     + P_{-} \left(d_1 w_0^{t-1} + d_2 e^{-\alpha(t-1)} \right) \right],
\end{eqnarray}
where $\alpha$ and the coefficients of each term are explicitly given
in the appendix.
By neglecting $e^{-\alpha N/2}$ order terms together with $w_0^{N/2}$
we can make use of the formula similar to \eqn{eqn:formula}.
Finally the effective vertex becomes
\begin{eqnarray}
\frac{T_A^{(1)}}{1-w_0^2} \gamma_\mu \gamma_5
&=&
\left( w_0^{(N-s)} P_+ + w_0^{(s-1)} P_- \right)
%\left( \Gamma^{(0)}_\mu \gamma_5 + \Gamma^{(1)}_\mu \gamma_5
%- \gamma_5 \Gamma^{(2)}_\mu + \Gamma^{(3)}_\mu \gamma_5 \right)(s,t)
\left( \Gamma^{(0)}_\mu + \Gamma^{(1)}_\mu
 + \Gamma^{(2)}_\mu + \Gamma^{(3)}_\mu \right)(s,t) \gamma_5
\nn\\&&\times
\left( w_0^{(N-t)} P_- + w_0^{(t-1)} P_+ \right)
\label{eqn:TA2}
\end{eqnarray}
with
\begin{eqnarray}
&&
\Gamma^{(0)}_\mu(s,t)=\int_{-\pi}^\pi \frac{d^4 l}{(2\pi)^4} \sum_{\nu \rho}
V_{1\nu}^a(0,l) S_F(l)_{s,u} A^{(0)}_\mu(l) S_F(l)_{u,t}
V_{1\rho}^b(-l,0) G_{\nu \rho}^{ab}(l),
\label{eqn:verteces4}
\\&&
\Gamma^{(1)}_\mu(s,t)=\int_{-\pi}^\pi \frac{d^4 l}{(2\pi)^4} \sum_\nu
V_{1\nu}^a(0,l) S_F(l)_{s,t} A^{(1)b}_\mu(l) G_{\nu \mu}^{ab}(l),
\\&&
\Gamma^{(2)}_\mu(s,t)=\int_{-\pi}^\pi \frac{d^4 l}{(2\pi)^4} \sum_\nu
A^{(1)a}_\mu(l) S_F(l)_{s,t} V_{1\nu}^b(-l,0) G_{\mu \nu}^{ab}(l),
\\&&
\Gamma^{(3)}_\mu(s,t)=\int_{-\pi}^\pi \frac{d^4 l}{(2\pi)^4}
A^{(2)ab}_\mu G_{\mu \mu}^{ab}(l) \delta(s,t).
\label{eqn:verteces7}
\end{eqnarray}
Here we have used the fact that $\Gamma_\mu^{(i)}$ are proportional to
$\gamma_\mu$ and therefore anticommute with $\gamma_5$.

The half-circle and tadpole 
diagrams in Fig.~\ref{fig:wave} contribute to the wave function
renormalization,
\begin{eqnarray}
&&
%\nabla_\mu
\vev{A_\mu^a(q) q(k) \bq(p)}_1^{\rm wave}
\nn\\&&=
%i(-k_\mu+p_\mu) \Biggl\{
\frac{1-w_0^2}{i\kslash +(1-w_0^2)m}
\left( w_0^{(N-s)} P_+ + w_0^{(s-1)} P_- \right)
\gamma_\mu \epsilon(s,t) S_F(p)_{t,t'} \Sigma(p,m)_{t',u}
\nn\\&&\times
\Biggl[
 \left( w_0^{(N-u)} P_- + w_0^{(u-1)} P_+ \right)
-\left( w_0^{(u-1)} P_- + w_0^{(N-u)} P_+ \right)
\frac{w_0}{1-w_0^2} i\pslash \Biggr]
\frac{1-w_0^2}{i\pslash + (1-w_0^2) m}
\nn\\&&+
\frac{1-w_0^2}{i\kslash + (1-w_0^2) m}
\Biggl[
 \left( w_0^{(N-s)} P_+ + w_0^{(s-1)} P_- \right)
-\frac{w_0}{1-w_0^2} i\kslash 
 \left( w_0^{(s-1)} P_+ + w_0^{(N-s)} P_- \right) \Biggr]
\nn\\&&\times
\Sigma(k,m)_{s,t} S_F(k)_{t,t'} \gamma_\mu \epsilon(t',u)
\left( w_0^{(N-u)} P_- + w_0^{(u-1)} P_+ \right)
\frac{1-w_0^2}{i\pslash +(1-w_0^2)m}.
%\Biggr\}.
\end{eqnarray}
We set the external quark momentum and mass to physical scale and extract
the leading order terms in these variables.
The fermion self-energy is the same as that is given in our previous paper
\cite{AIKT98,Aoki-Taniguchi}.
The difference is the existence of the fermion propagator in the internal
line, which is expanded as follows together with the external line damping
factor,
\begin{eqnarray}
&&
\left( w_0^{(N-s)} P_- + w_0^{(s-1)} P_+ \right) S_F(p)_{s,t}
\nn\\&&\to
\frac{1}{i\pslash+(1-w_0^2) m}
\left[
   \left( w_0^{N-t} P_+ + w_0^{t-1} P_- \right)
+m w_0 \left( w_0^{t-1} P_+ + w_0^{N-t} P_- \right)
\right],
\\&&
S_F(p)_{t,s} \left( w_0^{(N-s)} P_+ + w_0^{(s-1)} P_- \right)
\\&&\to
\left[
   \left( w_0^{t-1} P_+ + w_0^{N-t} P_- \right)
+m w_0 \left( w_0^{N-t} P_+ + w_0^{t-1} P_- \right)
\right]
\frac{1}{i\pslash+(1-w_0^2) m}.
\end{eqnarray}
Here we notice that the leading and the next to leading order terms give
the same order contribution.

With this expansion we have
\begin{eqnarray}
&&
\vev{A_\mu^a(q) q(k) \bq(p)}_1^{\rm wave}
\nn\\&&=
\frac{1}{i\kslash +(1-w_0^2)m}
\gamma_\mu \gamma_5
\frac{1-w_0^2}{i\pslash+(1-w_0^2) m}
\Sigma_v(p,m)
\frac{1-w_0^2}{i\pslash + (1-w_0^2) m}
\nn\\&&+
\frac{1-w_0^2}{i\kslash +(1-w_0^2)m}
\Sigma_v(k,m)
\frac{1-w_0^2}{i\kslash+(1-w_0^2) m}
\gamma_\mu \gamma_5
\frac{1}{i\pslash + (1-w_0^2) m},
\end{eqnarray}
where $\Sigma_v$ is defined by
\begin{eqnarray}
\Sigma_v(p,m) &=&
\left[
   \left( w_0^{N-s} P_+ + w_0^{s-1} P_- \right)
+m w_0\left( w_0^{s-1} P_+ + w_0^{N-s} P_- \right)
\right]
\Sigma(p,m)_{s,t}
\nn\\&&\times
\Biggl[
 \left( w_0^{N-t} P_- + w_0^{t-1} P_+ \right)
-\left( w_0^{t-1} P_- + w_0^{N-t} P_+ \right)
\frac{w_0}{1-w_0^2} i\pslash \Biggr].
\end{eqnarray}
The quark self-energy $\Sigma_{st}$ is expanded in terms of $p_\mu$ and $m$,
\begin{equation}
\Sigma(p,m)_{st} = \Sigma(0)_{st}
+\frac{\partial\Sigma(0)_{st}}{\partial i p_\mu} i p_\mu
+\frac{\partial\Sigma(0)_{st}}{\partial m} m
+O(p^2,m^2,pm),
\end{equation}
and $\Sigma_v$ becomes
\begin{eqnarray}
\Sigma_v(p,m) &=&
\frac{1}{1-w_0^2} i\pslash \left( Z_2^{(1)} - \frac{1}{2} Z_w^{(1)} \right)
+ m \left(Z_m^{(1)} + \frac{1}{2} Z_w^{(1)} \right),
\end{eqnarray}
where
\begin{eqnarray}
&&
\frac{1}{1-w_0^2} Z_2^{(1)} =
\left( w_0^{N-s} P_- + w_0^{s-1} P_+ \right)
\frac{1}{4} {\rm tr} \left( \gamma_\mu 
\frac{\partial\Sigma(0)_{st}}{\partial i p_\mu}
\right)
\left( w_0^{N-t} P_- + w_0^{t-1} P_+ \right),
\\&&
Z_m^{(1)} =
\left( w_0^{N-s} P_+ + w_0^{s-1} P_- \right)
\frac{\partial\Sigma(0)_{st}}{\partial m}
\left( w_0^{N-t} P_- + w_0^{t-1} P_+ \right),
\\&&
Z_w^{(1)} =
2 w_0 \left( w_0^{N-s} P_- + w_0^{s-1} P_+ \right)
\Sigma(0)_{st}
\left( w_0^{t-1} P_+ + w_0^{N-t} P_- \right).
\end{eqnarray}

Summing up all the contribution we have one loop level full Green's
function,
\begin{eqnarray}
&&
%\nabla_\mu
 \vev{A_\mu(n) q(y) \bq(z)}_1^{\rm full}
=\vev{A_\mu(n) q(y) \bq(z)}_0
+\vev{A_\mu(n) q(y) \bq(z)}_1^{\rm vertex}
+\vev{A_\mu(n) q(y) \bq(z)}_1^{\rm wave}
\nn\\&&=
%i(-k_\mu+p_\mu)
\frac{(1-w_0^2) (Z_w Z_2)^{\frac{1}{2}}}
{i\kslash+(1-w_0^2) Z_w Z_m^{-1} m}
\frac{Z_A}{1-w_0^2} \gamma_\mu \gamma_5
\frac{(1-w_0^2) (Z_w Z_2)^{\frac{1}{2}}}
%\frac{(1-w_0^2) (Z_w Z_2)^{\frac{1}{2}} Z_2}
{i\pslash+(1-w_0^2) Z_w Z_m^{-1} m},
\end{eqnarray}
where
\begin{eqnarray}
&&
Z_A = 1+Z_2^{(1)}+T_A^{(1)},
\\&&
Z_2 = 1+Z_2^{(1)},
\\&&
Z_w = 1-Z_w^{(1)},
\\&&
Z_m^{-1} = 1-Z_m^{(1)}+Z_2^{(1)}.
\end{eqnarray}

In the following we will show the relation
\begin{eqnarray}
\frac{\partial\Sigma(0)_{st}}{\partial i p_\mu}
=- \left( \Gamma^{(0)}_\mu + \Gamma^{(1)}_\mu
 + \Gamma^{(2)}_\mu + \Gamma^{(3)}_\mu \right)(s,t),
\label{eqn:deriv}
\end{eqnarray}
and consequently  $Z_2^{(1)}=-T_A^{(1)}$.
We start by writing $\Sigma(p,m)$ explicitly
\begin{eqnarray}
\Sigma(p,m)_{st}
&=&
\int_{-\pi}^\pi \frac{d^4 l}{(2 \pi)^4}
\sum_{\nu \rho}
V_{1\nu}^a(-p,l+p) S_{\rm F} (p+l)_{s,t} V_{1\rho}^b(-l-p,p)
G_{\nu \rho}^{ab}(l)
\nn\\
&+& 
\int_{-\pi}^\pi \frac{d^4 l}{(2 \pi)^4}
V_{2\nu\nu}^{ab}(-p,p) G_{\nu \nu}^{ab}(l) \delta_{s,t}.
\end{eqnarray}
A derivative with $p_\mu$ is given by
\begin{eqnarray}
\frac{\partial\Sigma(0)_{st}}{\partial i p_\mu}
=\left( I^{(0)}_\mu + I^{(1)}_\mu + I^{(2)}_\mu + I^{(3)}_\mu \right)(s,t),
\end{eqnarray}
where
\begin{eqnarray}
&&
I^{(0)}_\mu =
\int_{-\pi}^\pi \frac{d^4 l}{(2 \pi)^4}
\sum_{\nu \rho}
V_{1\nu}^a(0,l)
\left.\frac{\p S_{\rm F}(p+l)_{s,t}}{\p ip_\mu}\right|_{p_\mu=0}
V_{1\rho}^b(-l,0)
G_{\nu \rho}^{ab}(l)
= -\Gamma_\mu^{(0)}(s,t),
\label{eqn:I0}
\\&&
I^{(1)}_\mu =
\int_{-\pi}^\pi \frac{d^4 l}{(2 \pi)^4}
\sum_{\nu \rho}
\left. \frac{\p V_{1\nu}^a(-p,l+p)}{\p ip_\mu}\right|_{p_\mu=0}
 S_{\rm F} (l)_{s,t} V_{1\rho}^b(-l,0)
G_{\nu \rho}^{ab}(l)
= -\Gamma_\mu^{(1)}(s,t),
\\&&
I^{(2)}_\mu =
\int_{-\pi}^\pi \frac{d^4 l}{(2 \pi)^4}
\sum_{\nu \rho}
V_{1\nu}^a(0,l) S_{\rm F} (l)_{s,t}
\left. \frac{V_{1\rho}^b(-l-p,p)}{\p ip_\mu}\right|_{p_\mu=0}
G_{\nu \rho}^{ab}(l)
= -\Gamma_\mu^{(2)}(s,t),
\\&&
I^{(3)}_\mu =
\int_{-\pi}^\pi \frac{d^4 l}{(2 \pi)^4}
\left.\frac{\p V_{2\nu\nu}^{ab}(-p,p)}{\p ip_\mu}\right|_{p_\mu=0}
G_{\nu \nu}^{ab}(l) \delta_{s,t}
= -\Gamma_\mu^{(3)}(s,t).
\label{eqn:I3}
\end{eqnarray}
For the second equality in each equation we use the relation
\begin{eqnarray}
\left.\frac{\p S_{\rm F}(p+l)_{s,t}}{\p ip_\mu}\right|_{p_\mu=0}
%&=&-S_F(l)_{su}\left(\gamma_\mu \cos l_\mu -ir \sin l_\mu\right)S_F(l)_{ut},
&=& -S_F(l)_{su} A_\mu{(0)} S_F(l)_{ut},
\label{eqn:def-prop}
\\
\left. \frac{\p V_{1\nu}^a(-p,l+p)_{s,t}}{\p ip_\mu}\right|_{p_\mu=0}
&=& -A_\mu^{(1)a}(l)\delta_{\mu\nu}\delta_{s,t},
\\
\left. \frac{V_{1\rho}^b(-l-p,p)_{s,t}}{\p ip_\mu}\right|_{p_\mu=0}
&=& -A_\mu^{(1)b}(l)\delta_{\mu\rho}\delta_{s,t},
\\
\left. \frac{V_{2\nu\nu}^{ab}(-p,p)_{s,t}}{\p ip_\mu}\right|_{p_\mu=0}
&=& -\frac{1}{2}\left(A_\mu^{(2)ab} +A_\mu^{(2)ba}\right)\delta_{\mu\nu}
\delta_{s,t}.
\end{eqnarray}
The Eq.~\eqn{eqn:def-prop} is given by differentiating the definition of
the propagator
\begin{eqnarray}
\left[ i\gamma_\mu \sin p_\mu +W^+(p) P_+ +W^-(p) P_- \right] S_F(p) = 1
\end{eqnarray}
and multiplying $S_F(p)$ from the left.

Equalities in \eqn{eqn:I0} to \eqn{eqn:I3} prove eq.~\eqn{eqn:deriv},
which leads to  $Z_A=1$. This 
means that the axial vector current does not receive renormalization.

The one loop correction to the explicit breaking term is given by the
first diagram in Fig.~\ref{fig:vertex} and those in Fig.~\ref{fig:wave}
whose current vertex is now replaced by
\begin{eqnarray}
X^a_{st} = \left( P_+ \delta_{s,N/2} \delta_{t,N/2+1}
- P_- \delta_{s,N/2+1} \delta_{t,N/2} \right) \lambda^a.
\end{eqnarray}
In the diagram of Fig.~\ref{fig:wave} the vertex $X^a$ is directly
multiplied by the damping factor in external quark line and gives the
suppression factor $w_0^{N/2}$.
A contributions form the remaining part of the diagram is ${\cal O}(1)$ at
most and  the total contribution, which is a product of the two,
vanish in $N\to\infty$ limit.
For the first diagram in Fig.~\ref{fig:vertex} the damping factor 
appears in the contracted fermion propagator,
whose explicit form is given in the appendix.
Since the vertex $X^a$ is localized around $s,t\simeq N/2$,
the damping factor produces suppressions
such as $w_0^{N/2}$ or $e^{-\alpha N/2}$,
and therefore the contribution from Fig.~\ref{fig:vertex}
again vanishes in the large $N$ limit.

Consequently the renormalized axial Ward-Takahashi identity is satisfied
without fine tuning, 
\begin{eqnarray}
0&=&
i(-k_\mu+p_\mu)
\frac{1}{i\kslash+(1-w_0^2) Z_w Z_m^{-1} m} \lambda^a
Z_A \gamma_\mu \gamma_5
\frac{(1-w_0^2) Z_w Z_2}{i\pslash+(1-w_0^2) Z_w Z_m^{-1} m}
\nn\\&-&
2 \frac{1}{i\kslash + (1-w_0^2)Z_w Z_m^{-1} m} \lambda^a
\left(1-w_0^2\right) Z_w Z_P m \gamma_5
\frac{\left(1-w_0^2\right) Z_w Z_2}{i\pslash + (1-w_0^2)Z_w Z_m^{-1} m}
\nn\\&+&
\lambda^a \gamma_5
\frac{(1-w_0^2)Z_w Z_2}{i\pslash + (1-w_0^2)Z_w Z_m^{-1} m}
+
\frac{(1-w_0^2)Z_w Z_2}{i\kslash + (1-w_0^2)Z_w Z_m^{-1} m}
\gamma_5 \lambda^a,
\end{eqnarray}
together with conditions such that $Z_A=1$ and $Z_P=Z_m^{-1}$\cite{AIKT98}.

\section{Conclusion}

In this paper we have calculated the one loop correction to the axial
Ward-Takahashi identity and a renormalization factor for
the ``conserved'' axial vector current in domain-wall QCD.
Starting from the Green's function with the operators constructed from 
the physical quark field,
we find that the axial Ward-Takahashi identity holds exactly without
fine tuning at this order of the perturbation theory, and
consequently the axial vector current defined by Furman and Shamir 
receives no renormalization, $Z_A=1$.
As discussed in Ref.~\cite{Shamir95}, one may expect that
this property holds in all orders of the perturbation theory
as long as zero modes exist.
So one should use this almost conserved axial current to
extract the pion decay constant in numerical simulations.

\section*{Acknowledgements}

This work is supported in part by the Grants-in-Aid for
Scientific Research from the Ministry of Education, Science and Culture
(No.~2373). 
Y.~T. is supported by Japan Society for Promotion of Science.

% ------------------------ appendix ----------------------
\section*{Appendix A. Feynman Rules}

The fermion propagator is defined by
\begin{eqnarray}
S_F(p)_{st} =
\left( -i\gamma_\mu \sin p_\mu + W^- \right)_{su} G_R(u,t) P_+
+
\left( -i\gamma_\mu \sin p_\mu + W^+ \right)_{su} G_L(u,t) P_-
\end{eqnarray}
with
\begin{eqnarray}
W^{+}_{s,t} &=&
\pmatrix{
-W & 1  &        &    \cr
   & -W & \ddots &    \cr
   &    & \ddots & 1  \cr
 m &    &        & -W \cr
},\quad
W^{-}_{s,t} =
\pmatrix{
-W &        &        &  m \cr
1  & -W     &        &    \cr
   & \ddots & \ddots &    \cr
   &        & 1      & -W \cr
},
\\
W &=& 1-M -r \sum_\mu (1-\cos p_\mu),
\\
G_{R} (s, t) 
&=&
\frac{A}{F}
\Bigl[
-(1-m^2) \left(1-W e^{-\alpha}\right) e^{\alpha (-2N+s+t)}
-(1-m^2) \left(1-W e^\alpha\right) e^{-\alpha (s+t)}
\nn\\&&
-2W \sinh (\alpha) m
 \left( e^{\alpha (-N+s-t)} + e^{\alpha (-N-s+t)} \right)
\Bigr]
+ A e^{-\alpha |s-t|},
\\
G_{L} (s, t) 
&=&
\frac{A}{F}
\Bigl[
-(1-m^2) \left(1-We^{\alpha}\right) e^{\alpha (-2N+s+t-2)}
-(1-m^2) \left(1-We^{-\alpha}\right) e^{\alpha (-s-t+2)}
\nn\\&&
-2W \sinh (\alpha) m
 \left( e^{\alpha (-N+s-t)} +  e^{\alpha (-N-s+t)} \right)
\Bigr]
+ A e^{-\alpha |s-t|},
\\
\cosh (\alpha) &=& \frac{1+W^2+\sum_\mu \sin^2 p_\mu}{2W},
\\
A&=& \frac{1}{2W \sinh (\alpha)},
\\
F &=& 1-e^{\alpha} W-m^2 \left(1-W e^{-\alpha}\right).
\end{eqnarray}
The physical quark propagator is given by
\begin{eqnarray}
\vev{q(-p) \ovl{q}(p)} = 
 \frac{-i\gamma_\mu \sin p_\mu + \left(1-W e^{-\alpha}\right) m}
{-\left(1-e^{\alpha}W\right) + m^2 (1-W e^{-\alpha})}.
\label{eqn:phys-prop}
\end{eqnarray}

The gluon propagator can be written as
\begin{eqnarray}
G_{\mu \nu}^{ab} (p)
=\frac{1}{4\sin^2 p/2}
\left[\delta_{\mu \nu}
- (1-\alpha) \frac{4 \sin {p}_\mu/2 \sin {p}_\nu/2}{4 \sin^2 p/2}
\right]
 \delta_{ab},
\end{eqnarray}
where $\sin^2 p/2 = \sum_\mu \sin^2 p_\mu/2$.
The fermion-gluon interaction vertices which are
relevant for the one loop calculation are given by
\begin{eqnarray}
V_{1\mu}^a (k,p)_{st}
&=& -i g T^a \{ \gamma_\mu \cos \frac{1}{2}(-k_\mu + p_\mu)
  -i r \sin \frac{1}{2}(-k_\mu + p_\mu) \} \delta_{st},
\\
V_{2\mu\nu}^{ab} (k,p)_{st}
&=& \frac{1}{2} g^2 \frac{1}{2} \{T^{a}, T^{b}\}
\{ i \gamma_\mu \sin \frac{1}{2}(-k_\mu + p_\mu)
-r \cos \frac{1}{2} (-k_\mu + p_\mu) \}\delta_{\mu\nu} \delta_{st}.
\end{eqnarray}

The current vertices used in Eqs.~\eqn{eqn:vertices0} to
\eqn{eqn:vertices3} are given by expanding
the axial vector current in terms of the gauge field.
Only three kinds of them are relevant for one loop calculation,
\begin{eqnarray}
&&
A^{(0)}_\mu(l) = \gamma_\mu \cos l_\mu -ir \sin l_\mu,
\\&&
A^{(1)a}_\mu(l) =
-g \left( \gamma_\mu \sin \frac{l_\mu}{2}+ir \cos \frac{l_\mu}{2} \right) T^a,
\\&&
A^{(2)ab}_\mu = -\frac{g^2}{2} \gamma_\mu T^a T^b.
\end{eqnarray}

The explicit form of the contracted propagator with the damping factor
is given by
\begin{eqnarray}
&&
\left( w_0^{(N-s)} P_- + w_0^{(s-1)} P_+ \right) S_F(p)_{s,t}
\nn\\&&\quad=
-i\gamma_\mu \sin p_\mu
\left( w_0^{(N-s)} G_R(s,t) P_+ + w_0^{(s-1)} G_L(s,t) P_- \right)
\nn\\&&\quad
+\left( \left( w_0-W(p) \right) w_0^{s-1} G_R(s,t) +m G_R(N,t) \right) P_+
\nn\\&&\quad
+\left( \left( w_0-W(p) \right) w_0^{N-s} G_L(s,t) +m G_L(1,t) \right) P_-,
\\&&
S_F(p)_{t,s} \left( w_0^{(N-s)} P_+ + w_0^{(s-1)} P_- \right)
\nn\\&&\quad=
\left( P_{+} w_0^{(s-1)} G_L(s,t) + P_{-} w_0^{(N-s)} G_R(s,t) \right)
(-i \gamma_\mu \sin p_\mu)
\nn\\&&\quad
+P_{+} \left(w_0^{(N-s)} G_L(s,t) \left(w_0-W(p)\right) + m G_L(t,1) \right)
\nn\\&&\quad
+P_{-} \left(w_0^{(s-1)} G_R(s,t) \left(w_0-W(p)\right) + m G_R(t,N) \right),
\end{eqnarray}
where
\begin{eqnarray}
w_0^{s-1} G_R(s,t) &=&
\frac{A}{F}
\Bigl[
-(1-m^2) \left(1-W e^\alpha\right) e^{-2\alpha} \frac{1}{1-w_0 e^{-\alpha}}
e^{-\alpha (t-1)}
\nn\\&&
-2W \sinh (\alpha) m e^{-\alpha} \frac{1}{1-w_0 e^{-\alpha}}
e^{-\alpha (N-t)}
\Bigr]
\nn\\&&
+ A \frac{e^{-\alpha(t-1)}(1-w_0 e^{-\alpha})-2w_0^{(t-1)} w_0 \sinh \alpha}
{1+w_0^2-2w_0\cosh \alpha},
\\
w_0^{N-s} G_{R} (s, t) 
&=&
\frac{A}{F}
\Bigl[
-(1-m^2) \left(1-W e^{-\alpha}\right) \frac{1}{1-w_0 e^{-\alpha}}
e^{-\alpha (N-t)}
\nn\\&&
-2W \sinh (\alpha) m e^{-\alpha} \frac{1}{1-w_0 e^{-\alpha}}
e^{-\alpha (t-1)}
\Bigr]
\nn\\&&
+ A \frac{e^{-\alpha(N-t)}(1-w_0 e^{-\alpha})-2w_0^{(N-t)} w_0 \sinh \alpha}
{1+w_0^2-2w_0\cosh \alpha},
\\
w_0^{s-1} G_{L} (s, t) 
&=&
\frac{A}{F}
\Bigl[
-(1-m^2) \left(1-We^{-\alpha}\right) \frac{1}{1-w_0 e^{-\alpha}}
e^{-\alpha (t-1)}
\nn\\&&
-2W \sinh (\alpha) m e^{-\alpha} \frac{1}{1-w_0 e^{-\alpha}} e^{-\alpha(N-t)}
\Bigr]
\nn\\&&
+ A \frac{e^{-\alpha(t-1)}(1-w_0 e^{-\alpha})-2w_0^{(t-1)} w_0 \sinh \alpha}
{1+w_0^2-2w_0\cosh \alpha},
\\
w_0^{N-s} G_{L} (s, t) 
&=&
\frac{A}{F}
\Bigl[
-(1-m^2) \left(1-We^{\alpha}\right) e^{-2\alpha}  \frac{1}{1-w_0 e^{-\alpha}}
e^{-\alpha(N-t)}
\nn\\&&
-2W \sinh (\alpha) m e^{-\alpha} \frac{1}{1-w_0 e^{-\alpha}} e^{-\alpha(t-1)}
\Bigr]
\nn\\&&
+ A \frac{e^{-\alpha(N-t)}(1-w_0 e^{-\alpha})-2w_0^{(N-t)} w_0 \sinh \alpha}
{1+w_0^2-2w_0\cosh \alpha},
\\
G_{R} (N, t) 
&=&
\frac{A}{F}
\Bigl[
-(1-m^2) \left(1-W e^{-\alpha}\right) e^{-\alpha(N-t)}
\nn\\&&
-2W \sinh (\alpha) m e^{-\alpha} e^{-\alpha(t-1)}
\Bigr]
+ A e^{-\alpha (N-t)},
\\
G_{L} (1, t) 
&=&
\frac{A}{F}
\Bigl[
-(1-m^2) \left(1-We^{-\alpha}\right) e^{-\alpha (t-1)}
\nn\\&&
-2W \sinh (\alpha) m e^{-\alpha} e^{-\alpha (N-t)}
\Bigr]
+ A e^{-\alpha (t-1)}.
\end{eqnarray}

%%%%%%%%%%%%%%%%%%%%%%%%%%%%%%%%%%%%%%%%%%%%%%%%%%%%%%%%%%%%%%%%%%%%%%
\newcommand{\J}[4]{{\it #1} {\bf #2} (19#3) #4}
\newcommand{\MPL}{Mod.~Phys.~Lett.}
\newcommand{\IJMP}{Int.~J.~Mod.~Phys.}
\newcommand{\NP}{Nucl.~Phys.}
\newcommand{\PL}{Phys.~Lett.}
\newcommand{\PR}{Phys.~Rev.}
\newcommand{\PRL}{Phys.~Rev.~Lett.}
\newcommand{\AP}{Ann.~Phys.}
\newcommand{\CMP}{Commun.~Math.~Phys.}
\newcommand{\PTP}{Prog. Theor. Phys.}
\newcommand{\Suppl}{Prog. Theor. Phys. Suppl.}
%%%%%%%%%%%%%%%%%%%%%%%%%%%%%%%%%%%%%%%%%%%%%%%%%%%%%%%%%%%%%%%%%%%%%%

%%%%%%%%%%%%%%%%%%%%%%%%%%%%%%%%%%%%%%%%%%%%%%%%%%%%%%%%%%%%%%%%%%%%%

% Fig. 1

% Fig. 2
\begin{figure}
\begin{center}\begin{picture}(200,100)(0,0)
\ArrowLine(180,10)(160,30)
\ArrowLine(160,30)(100,90)
\ArrowLine(100,90)(40,30)
\ArrowLine(40,30)(20,10)
\Gluon(40,30)(160,30){5}{5}
\Vertex(40,30){3}
\Vertex(160,30){3}
\BCirc(100,90){3}
\Line(98,88)(102,92)
\Line(98,92)(102,88)
\Text(150,60)[l]{$l+p$}\LongArrow(145,65)(135,75)
\Text(50,75)[l]{$l+k$}\LongArrow(55,65)(45,55)
\Text(95,20)[l]{$l$}\LongArrow(100,20)(110,20)
\Text(98,100)[c]{$A_0, X$}
\Text(195,20)[l]{$p$}\LongArrow(190,15)(180,25)
\Text(5,20)[r]{$k$}\LongArrow(20,25)(10,15)
\Text(15,0)[l]{$q$}
\Text(180,0)[l]{$\ovl{q}$}
\end{picture}
\end{center}
\begin{center}
\begin{picture}(200,100)(0,0)
\ArrowLine(180,10)(100,90)
\ArrowLine(100,90)(40,30)
\ArrowLine(40,30)(20,10)
\GlueArc(50,80)(50,-100,10){-5}{10}
\Vertex(40,30){3}
\BCirc(100,90){3}
\Line(98,88)(102,92)
\Line(98,92)(102,88)
\Text(90,30)[l]{$l$}\LongArrow(95,35)(105,45)
\Text(50,75)[l]{$l+k$}\LongArrow(55,65)(45,55)
\Text(95,100)[l]{$A_1$}
\Text(195,20)[l]{$p$}\LongArrow(190,15)(180,25)
\Text(5,20)[r]{$k$}\LongArrow(20,25)(10,15)
\Text(15,0)[l]{$q$}
\Text(180,0)[l]{$\ovl{q}$}
\end{picture}
\end{center}
\begin{center}
\begin{picture}(200,100)(0,0)
\ArrowLine(180,10)(160,30)
\ArrowLine(160,30)(100,90)
\ArrowLine(100,90)(20,10)
\GlueArc(150,80)(52,170,280){-5}{10}
\Vertex(160,30){3}
\BCirc(100,90){3}
\Line(98,88)(102,92)
\Line(98,92)(102,88)
\Text(150,60)[l]{$l+p$}\LongArrow(145,65)(135,75)
\Text(110,30)[l]{$l$}\LongArrow(95,45)(105,35)
\Text(95,100)[l]{$A_1$}
\Text(195,20)[l]{$p$}\LongArrow(190,15)(180,25)
\Text(5,20)[r]{$k$}\LongArrow(20,25)(10,15)
\Text(15,0)[l]{$q$}
\Text(180,0)[l]{$\ovl{q}$}
\end{picture}
\end{center}
\begin{center}
\begin{picture}(200,120)(0,0)
\ArrowLine(180,10)(100,90)
\ArrowLine(100,90)(20,10)
\GlueArc(100,110)(20,-90,270){-5}{10}
\BCirc(100,90){3}
\Line(98,88)(102,92)
\Line(98,92)(102,88)
\Text(95,75)[l]{$A_2$}
\Text(195,20)[l]{$p$}\LongArrow(190,15)(180,25)
\Text(5,20)[r]{$k$}\LongArrow(20,25)(10,15)
\Text(15,0)[l]{$q$}
\Text(180,0)[l]{$\ovl{q}$}
\end{picture}
\end{center}
\caption{One loop diagrams which contribute to the operator vertex.}
\label{fig:vertex}
\end{figure}
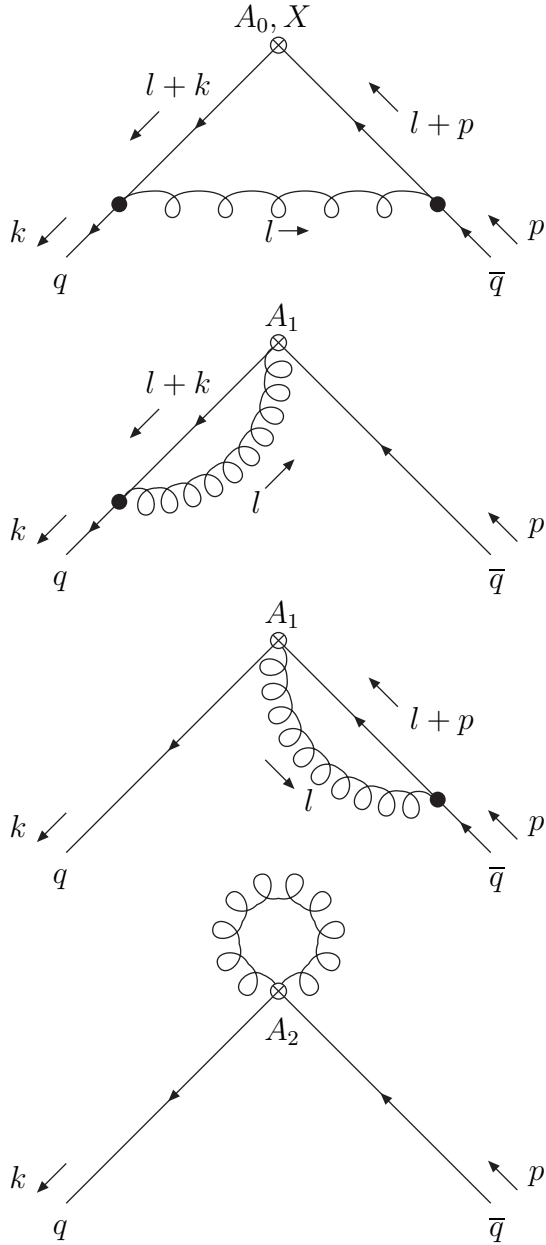

\end{document}